\documentclass[a4paper,11pt]{article}
\usepackage{pos}
\usepackage{natbib}
\usepackage{graphicx}
\usepackage{xspace}

\def\pp{\ensuremath{{\rm pp}}\xspace}
\def\ee{\ensuremath{e^+e^-}\xspace}
\def\ep{\ensuremath{ep}\xspace}
\def\pp{\ensuremath{pp}\xspace}

\def\ntrkoff{\ensuremath{{\rm N}_{\rm Trk}^{\rm Offline}}\xspace}

\def\dphi{\ensuremath{\Delta\phi}\xspace}

\title{Two-particle azimuthal correlations in \ee collisions at 91--209 GeV with archived ALEPH data at LEP-2}

\author*[a]{Yen-Jie Lee}
\author[a]{Yu-Chen Chen}
\author[a]{Yi Chen}
\author[b]{Paoti Chang}
\author[c]{Marcello Maggi}

\affiliation[a]{Massachusetts Institute of Technology, Cambridge, USA}

\affiliation[b]{National Taiwan University, Taipei, Taiwan}
\affiliation[c]{INFN Bari, Bari, Italy}

\emailAdd{yenjie@mit.edu}

\abstract{
We present the first measurement of two-particle angular correlations of charged particles produced in \ee annihilation up to $\sqrt{s}=$ 209 GeV. This analysis utilized the archived hadronic \ee data at center-of-mass energy between 91 and 209 GeV collected with the ALEPH detector at LEP between 1992 and 2000. The angular correlation functions are measured over a broad range of pseudorapidity and full azimuth as a function of charged particle multiplicity for the first time with LEP-2 data. At 91 GeV, no significant long-range correlation is observed in either the beam coordinate analysis or the thrust coordinate analysis, where the latter is sensitive to a medium expanding transverse to the color string between the outgoing $q\bar{q}$ pair from the Z boson decays. 
Results with \ee data at higher collision energy than 91 GeV, providing higher event multiplicity reach up to around 50, are presented for the first time. The thrust axis analysis shows a long-range near-side excess in the two-particle correlation function. We performed Fourier series decomposition of the two-particle correlation functions. In high multiplicity events with more than 50 particles, the extracted Fourier coefficients $v_2$ and $v_3$ magnitudes in data are larger than the MC reference.}

\FullConference{%
  41st International Conference on High Energy Physics - ICHEP2022\\
  6-13 July, 2022\\
  Bologna, Italy
}

\begin{document}
\maketitle

In nucleon-nucleon and heavy-ion collision experiments, two-particle angular correlations~\cite{STAR:2005ryu, STAR:2009ngv, PHOBOS:2009sau, Chatrchyan:2012wg, Aamodt:2011by, Adam:2019woz} are extracted for the study of the Quark-Gluon Plasma (QGP)~\cite{Busza:2018rrf} and the search for initial state correlation effects such as the Color Glass Condensate~\cite{Dumitru:2010iy}.
These measurements with various collision systems and collision energies have revealed a long-range angular correlation, the ridge-like structure~\cite{STAR:2009ngv, PHOBOS:2009sau}. Since the beginning of the LHC operation, the ridge structure has also been observed in high-multiplicity proton-proton collisions by the CMS collaboration~\cite{Khachatryan:2010gv} and confirmed by experiments at LHC and RHIC using smaller collision systems such as proton-proton~\cite{Aad:2015gqa}, proton-ion~\cite{CMS:2012qk, Abelev:2012ola, Aaij:2015qcq, ATLAS:2012cix}, and deuteron-ion~\cite{PHENIX:2013ktj} collisions. In heavy-ion collisions, the ridge structure is associated with the fluctuating initial state of the ions~\cite{Alver:2010gr}. However, the physical origin of the ridge structure in small systems is still under debate~\cite{Dusling:2013qoz, Bozek:2011if, He:2015hfa, Nagle:2018nvi}.

Recently, there has been growing interest in measuring two-particle correlations in even smaller collision systems of photonuclear PbPb~\cite{ATLAS:2021jhn, CMS:2022doq}, \ep~\cite{ZEUS:2019jya} and \ee~\cite{Badea:2019vey, Belle:2022fvl, The:2022lun}. They serve as counterparts complementary to the results in large collision systems and can be used to find the minimal conditions for collective behavior~\cite{Nagle:2017sjv}. Electron beams remove complications such as multiple parton interactions and initial state correlations. So far, no significant ridge-like signal has been observed in the most elementary electron-positron annihilation. Those measurements have provided additional insights on the ridge signal~\cite{Bierlich:2020naj, Castorina:2020iia, Agostini:2021xca, Larkoski:2021hee, Baty:2021ugw}.

In these proceedings, we report on the first measurement of two-particle angular correlation functions in high multiplicity \ee annihilation events at $\sqrt{s}= 91$--$209$ GeV with archived ALEPH data taken at LEP-2. Due to the higher collision energies above the $Z^0$ pole, the data set enables studies of higher multiplicity events compared to LEP-1 data. Moreover, the analyzed dataset will contain different underlying physics processes.


The analysis is performed using the data sample taken by the ALEPH detector between 1992 to 2000, corresponding to collision energies of $\sqrt{s} = 91$--209~GeV. Unlike the ``Z resonance'' sample taken at the 91.2 GeV, where $Z$-decays dominate over other processes, a wide variety of processes other than the $e^+e^- \to q\bar{q}$ fragmentation also make non-negligible contributions in the ``high energy sample'' with $\sqrt{s}$ between 183 and 209 GeV. The initial-state QED radiation is significant in high-energy events, which is an effect called ``radiative-return-to-$Z$.'' We follow the selection criteria studied by the ALEPH collaboration~\cite{ALEPH:2003obs}, to ensure the analyzed events are produced from high center-of-mass energy collisions.
QED ($e^+e^- \to \tau^+\tau^-$) and two-photon events contribute particularly in the low-multiplicity region. 
We follow the same hadronic event selection criteria as in the previous LEP-1 work~\cite{Badea:2019vey}, requiring the polar angle of the event sphericity axis to be within $7\pi/36$ to $29\pi/36$. Events having less than five tracks and the reconstruction energy smaller than $15$~GeV are rejected.

For high-energy events above the $W$-pair production threshold,  four-fermion processes mediated by the double- or single- $W$ or $Z$ boson are the sub-dominant contributions. 
The topology of four-fermion processes differs largely from that of the $e^+e^- \to q\bar{q}$; more importantly, there can be non-trivial final-state interconnection (FSI) effects such as Color Reconnection (CR) and Bose-Einstein Correlation (BEC) in hadronic decays of boson pairs. 
These physics processes might accommodate promising potentials for the ridge-signal search, which is beyond our current understanding of the \ee collision system. A recent theoretical work~\cite{Larkoski:2021hee} calculating by assuming the large-$N$ limit also makes remarks that for processes beyond the leading order $e^+e^- \to q\bar{q}$  events via $Z$ decays, there is no suppression for generating such collectivity correlations.

We utilize the Monte Carlo (MC) events simulated by ALEPH collaboration to study the reconstruction effects and correct the data. Archived $\textsc{pythia}$ 6.1~\cite{Sjostrand:2000wi} MC simulation samples, which were produced with the years 1994 and 1997-2000 run detector condition by the ALEPH collaboration, were the available archived MC sample at the time of this analysis. The MC samples are used to derive tracking efficiency and event selection corrections, and different MC processes are weighted according to the calculated cross-sections.

\begin{figure}[t]
\centering
\includegraphics[width=0.42\textwidth]{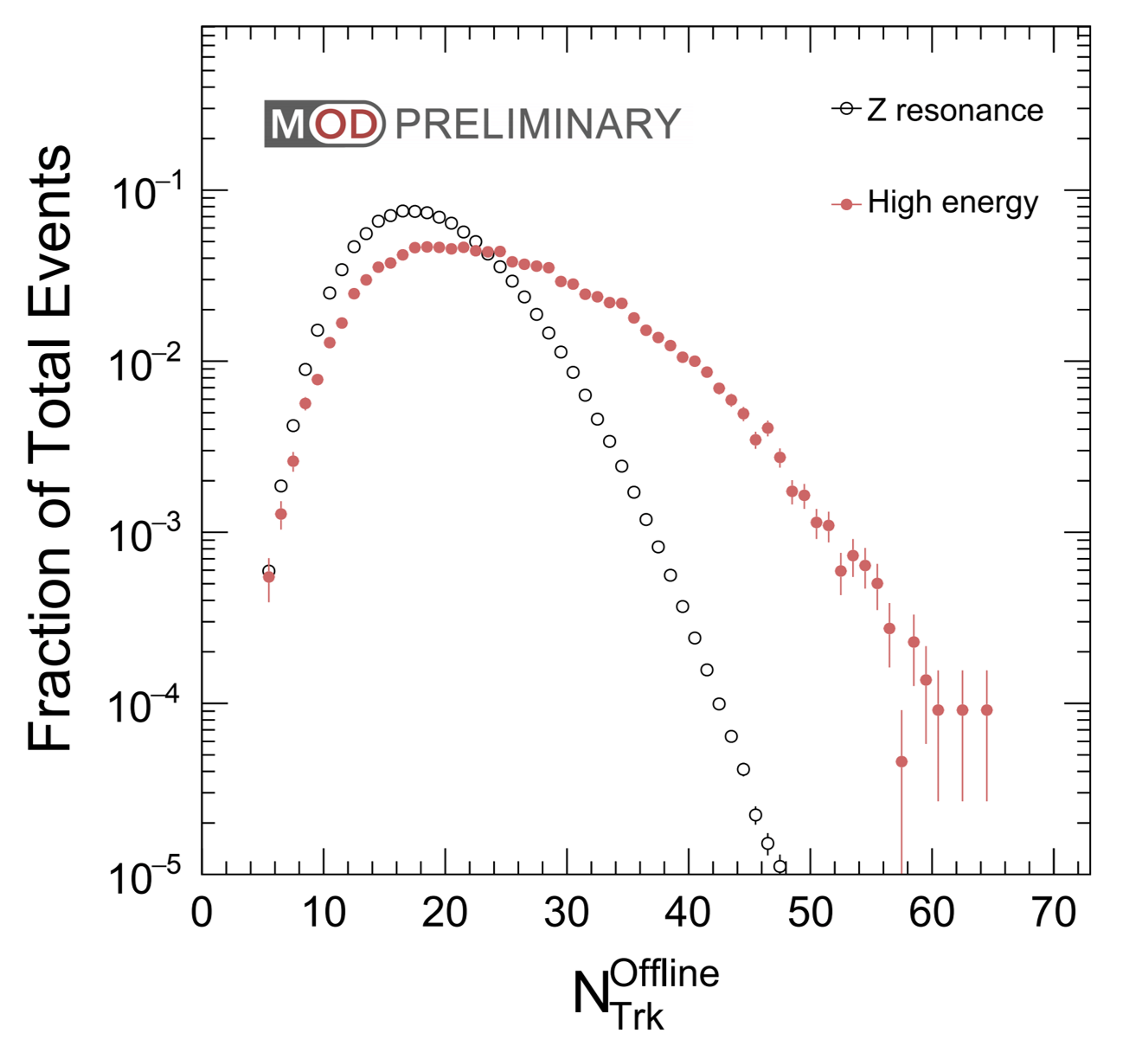}
\includegraphics[width=0.40\textwidth]{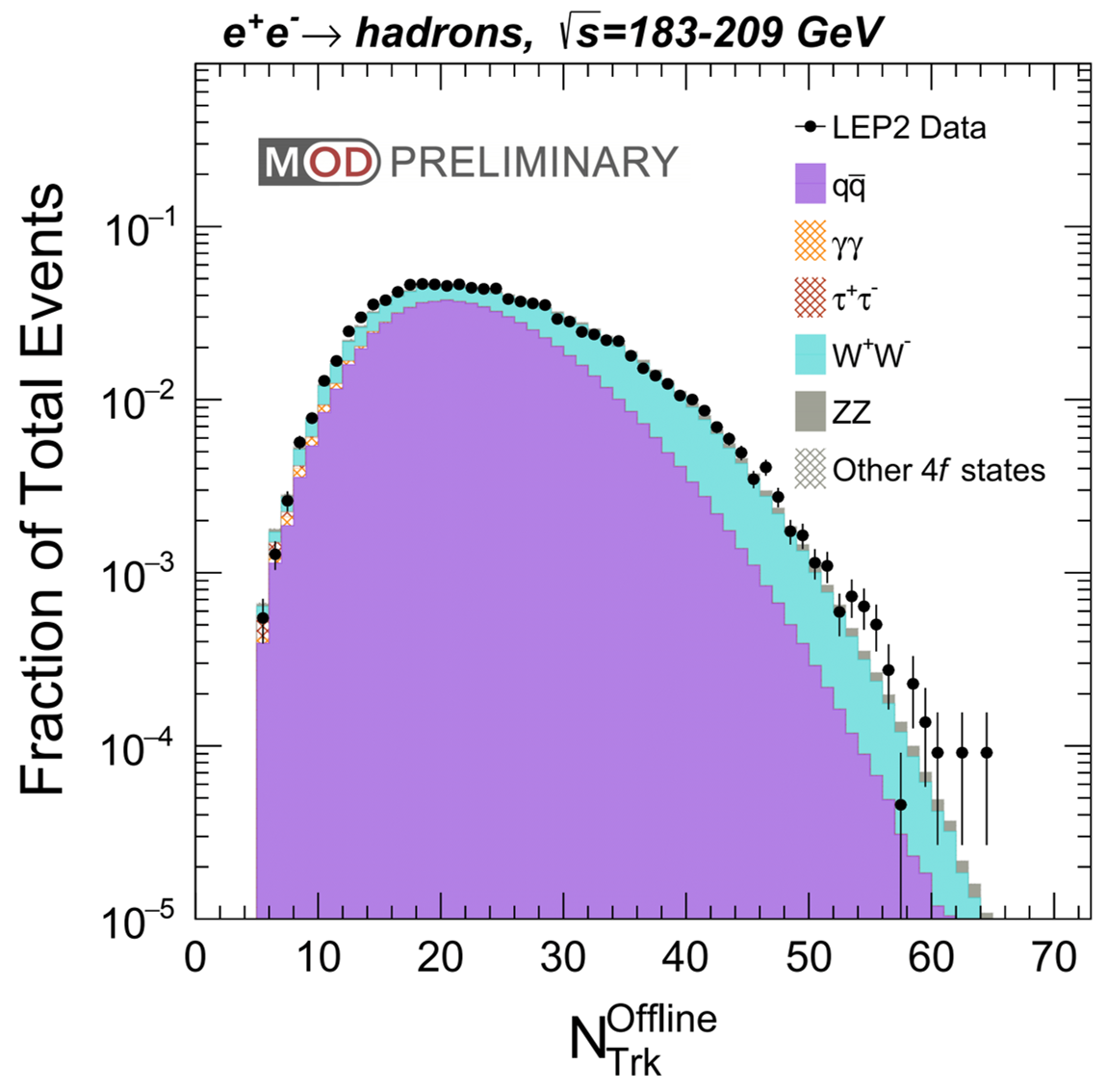}
\caption{(Left panel) Charged particle multiplicity distributions from the Z pole and high energy samples (Right panel) Multiplicity distributions in data compared to MC simulation for the high-energy sample
}
\label{fig:NtrkOffline}      
\end{figure}

Charged particle candidates must have at least four TPC hits and restrict from the interaction point within a loose allowed distance, where the radial displacement $d_0<2$~cm and the longitudinal displacement $z_0<10$~cm. 
The transverse momentum of a track should be greater than 0.2~GeV and with the absolute cosine of the polar angle smaller than 0.94. After all selections, the subsequent offline multiplicity (\ntrkoff) is used for studying the correlation function's multiplicity dependence.
Figure~\ref{fig:NtrkOffline} shows the offline multiplicity distributions in data collected at different collision energies. The high-energy sample provides a significantly higher multiplicity reach compared to the Z resonance sample. The MC simulation provides an excellent description of the data. While the $q\bar{q}$ final states dominant the lower multiplicity range, $W^{+}W^{-}$ contribution increases with \ntrkoff and becomes the most significant contribution at high multiplicity.

Two-particle correlations are then calculated by following heavy ion collisions and hadron collisions measurements, which have been documented in ref.~\cite{Badea:2019vey, Belle:2022fvl,The:2022lun}. Following the previous studies in \ee systems, we analyze correlation functions with the ``thrust-axis'' coordinates. Thrust axis $\hat{n}$ is used as the $z$-axis, and the choice of $\phi=0$ (or new reference $x$-axis) is assigned with $\hat{n}\times(\hat{n}\times\hat{z})$.
Since the outgoing direction of each event is randomly orientated, a thrust-mixing reweighting correction on the event topology -- the pseudorapidity and the azimuthal angle distributions with respect to the thrust axis ($\eta_T, \phi_T$) -- is adopted to match the mixed events' fragmenting topology to the physical ones.

\begin{figure}[t]
\centering
\includegraphics[width=0.42\textwidth]{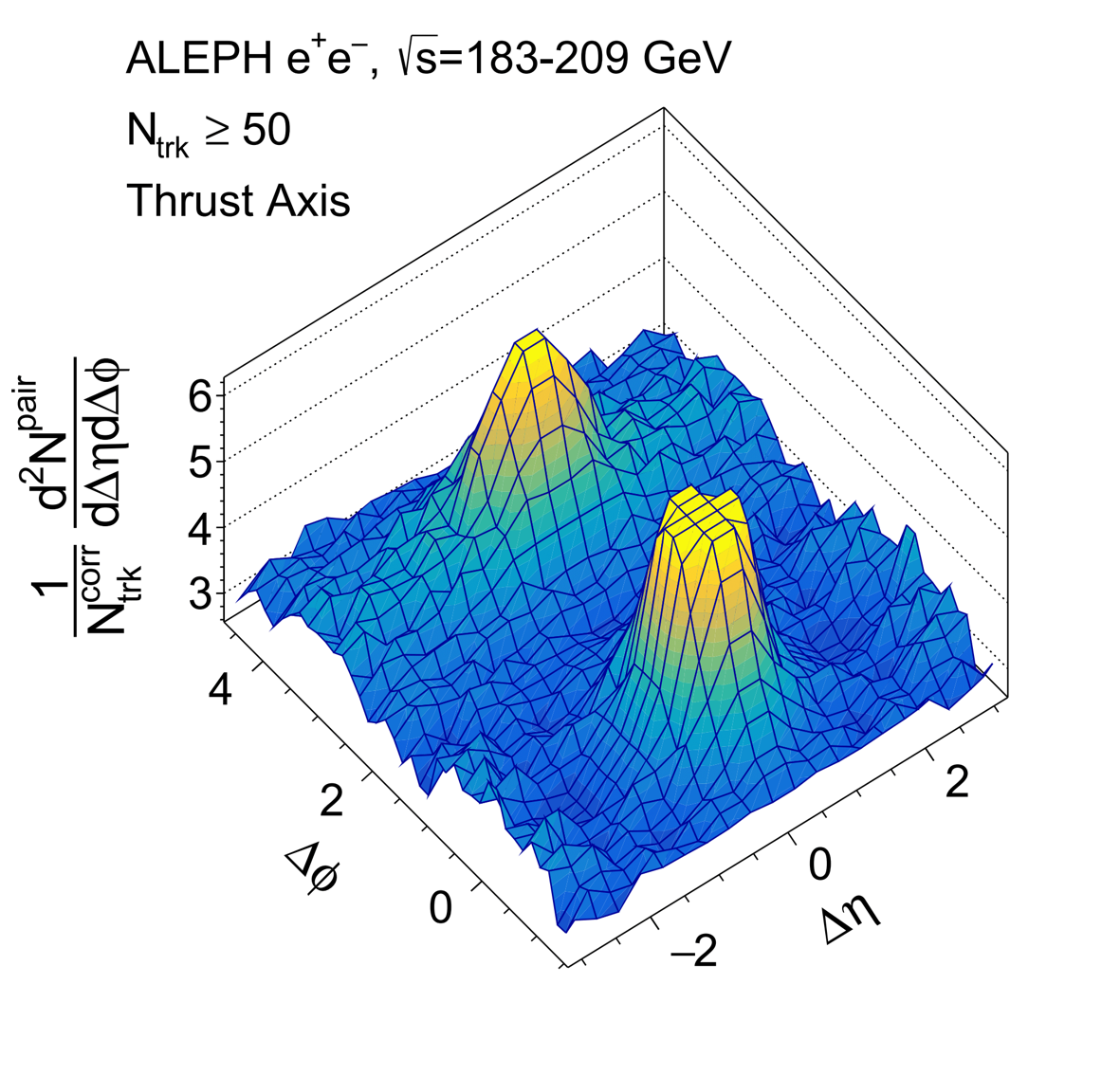}
\includegraphics[width=0.40\textwidth]{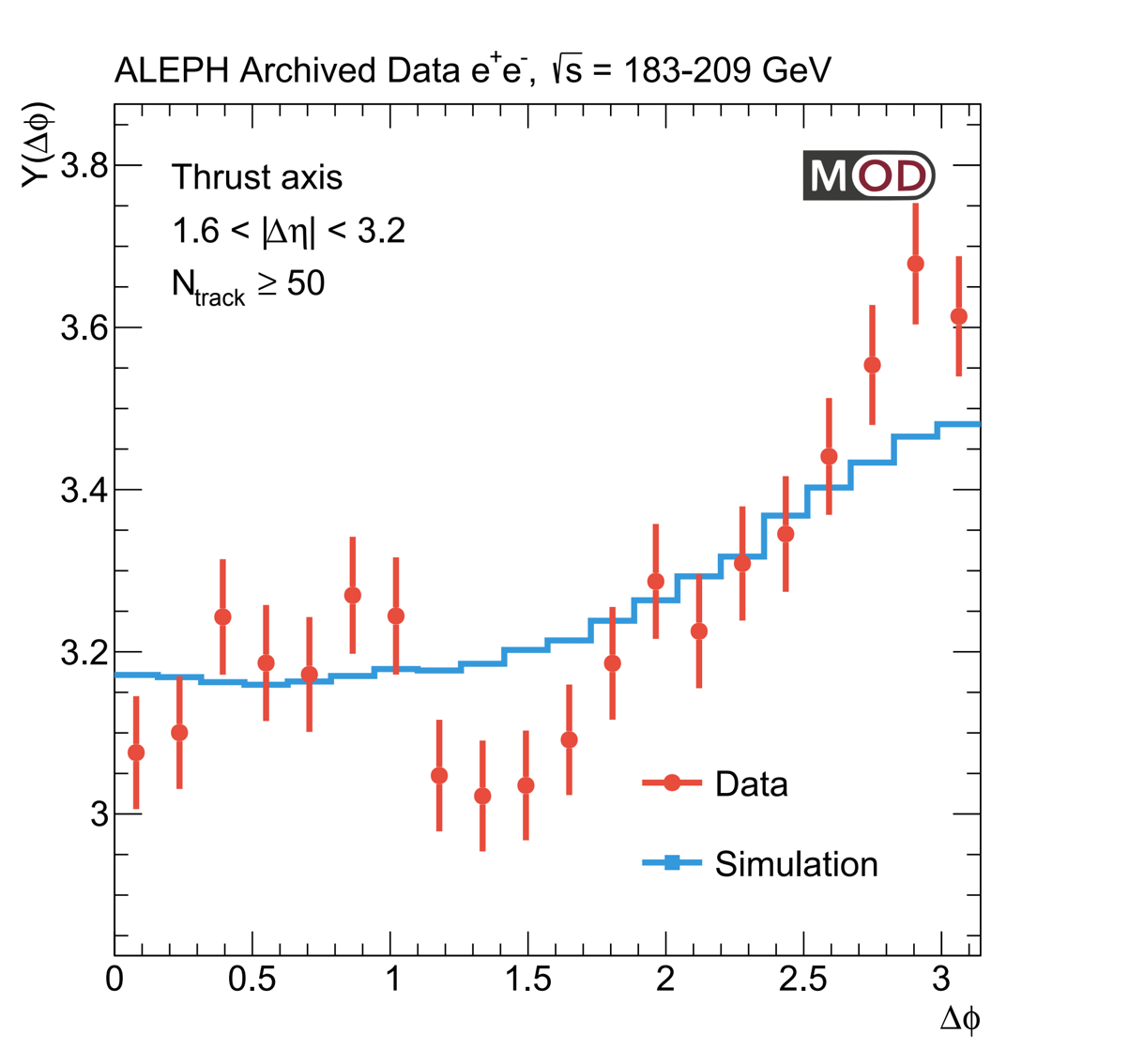}
        
\caption{(Left panel) Two-particle correlation function in events with $\ntrkoff> 50$. (Right panel) Associated yield as a function of azimuthal angle difference $|\Delta\phi|$}
\label{fig:correlation}      
\end{figure}

In this work, we study two-particle correlations in different multiplicity classes: [5,10), [10,20), [20,30), [30,40), [40,50), and [50,$\infty$). MC simulation describes the data in low multiplicity events with $\ntrkoff<40$. On the other hand, in the highest multiplicity class with $\ntrkoff>50$, a long-range near-side signal appears in data that is not described by MC simulation, which is shown in Figure~\ref{fig:correlation}.

\begin{figure}[t]
\centering
\includegraphics[width=0.43\textwidth]{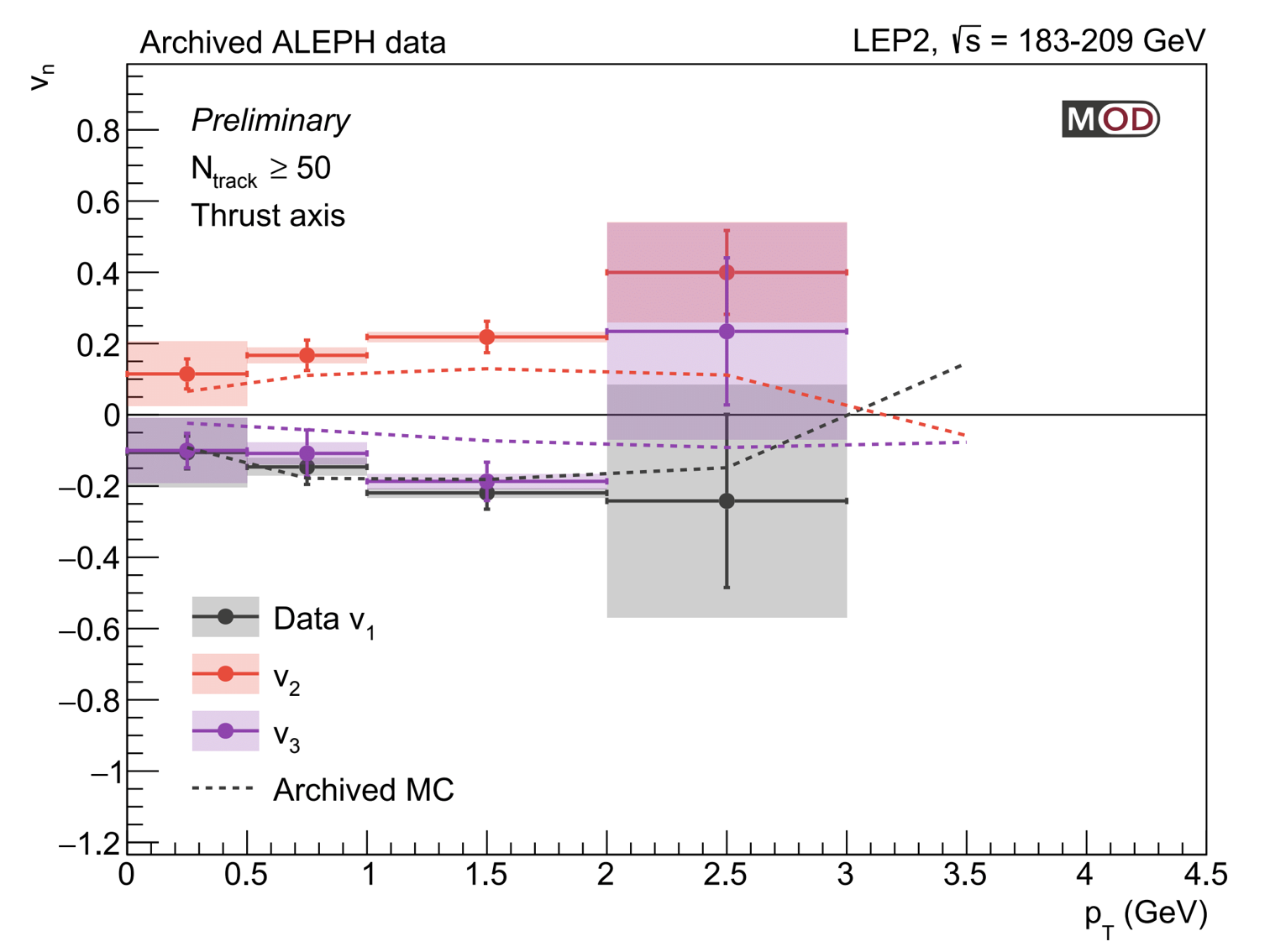}
\includegraphics[width=0.42\textwidth]{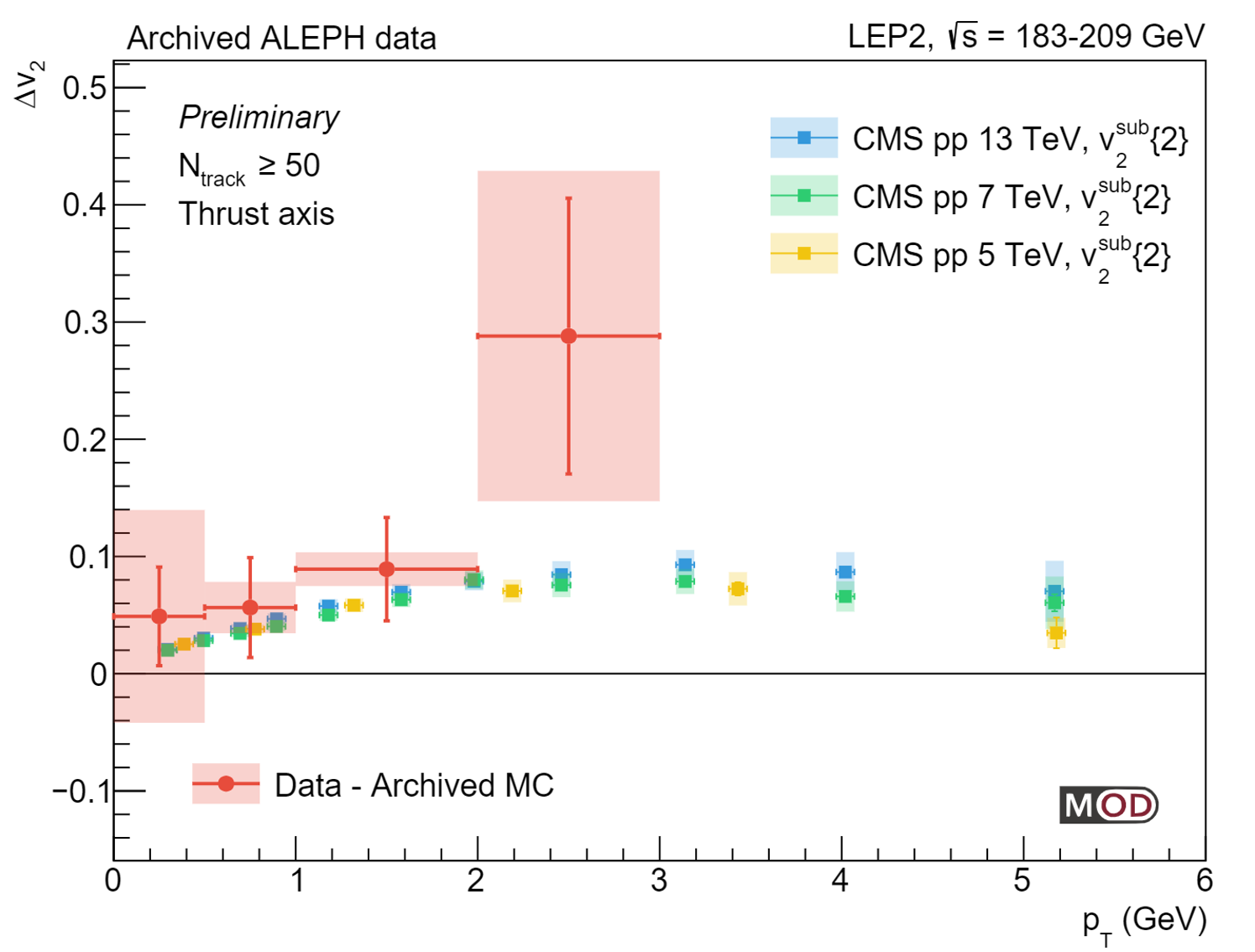}
        
\caption{(Left panel) The extracted $v_n$ in data (shown as dots) compared to archived MC (dash lines) (Right panel) $\Delta v_2$ (=$v_2 - v_2^{MC})$ compared to the extracted $v^{\rm sub}_2$ in \pp collisions
}
\label{fig:Dv2VsPt_beam}
\end{figure}

We also explored two-particle correlations with requirements on both particles' transverse momenta $p_T$. The trigger particle and the associated particle are therefore required to fall within the same $p_T$ class: $[0, 0.5)$, $[0.5, 1)$, $[1,2)$, $[2,3)$ and $[3, \infty)$ (in GeV). To quantify flow-like signatures, we follow the Fourier decomposition analysis~\cite{CMS:2011cqy, ALICE:2011svq, ATLAS:2012at}. 
The possible non-flow effects are suppressed in the large $|\Delta \eta|$ region.
In this analysis, the long-range azimuthal differential yields with $1.6 \le |\Delta \eta| < 3.2$ are fitted with the Fourier series: 
$Y(\Delta\phi) = \frac{1}{{\rm N}_{\rm trk}^{\rm corr}}\frac{d{\rm N}^{\rm pair}}{d\Delta\phi} = \frac{{\rm N}^{\rm assoc}}{2\pi} \bigg( 1 + \sum_{n=1}^{6} 2 V_{n\Delta} \cos(n\Delta\phi) \bigg)$, where ${\rm N}^{\rm assoc}$ is the number of associated track pairs in the $|\Delta \eta|$ region of interest and within the full $\Delta \phi$ range for a particular $p_T$ bin.
In this work, we consider the Fourier fit template up to the sixth order, and the fitting range is the full $\Delta \phi$ range ($0 \leq |\dphi| \leq \pi$).
The Fourier coefficients $V_{n\Delta}$ can be further factorized into the single-particle Fourier harmonics product, with the assumption that the azimuthal anisotropy results from the hydrodynamic flow effect. The single-particle harmonic coefficient $v_n$ is related to the Fourier coefficient $V_{n\Delta}$ by
$V_{n\Delta} (= v_n^{\rm trig} \times v_n^{\rm assoc}) = v_n^2$, and
$v_n = \frac{V_{n\Delta}}{\sqrt{|V_{n\Delta}|}}$ since the trigger particle and the associated particle are within the same $p_T$ bin. 
Figure~\ref{fig:Dv2VsPt_beam} shows the extracted $v_n$ in data compared to archived MC simulation. Compared to MC, a systematically larger $v_2$ and $v_3$ in data is present. To further suppress the residual non-flow effect, the difference between data $v_2$ and MC is also presented in Figure~\ref{fig:Dv2VsPt_beam}. The $\Delta v_2$ is compatible with the extracted $v^{\rm sub}_2$, the $v_2$ after subtracting the result in low multiplicity event to suppress the non-flow contribution, in \pp collisions~\cite{CMS:2016fnw}.

In summary, we present the first measurement of two-particle angular correlations of charged particles produced in \ee annihilation up to $\sqrt{s}=$ 209 GeV for the first time with archived ALEPH LEP-2 data. A long-range near-side excess in the correlation function is seen in the thrust axis analysis in \ee collisions at $\sqrt{s}=183$ to 209 GeV. The two-particle correlation functions are also decomposed using the Fourier series. The extracted Fourier coefficients $v_n$ from data are compared to archived MC. In high multiplicity events with more than 50 particles, the extracted $v_2$ and $v_3$ magnitude in data are larger than the MC reference. The difference between data and MC $v_2$ is found to be compatible with the measured $v_2^{\rm sub}$ in high multiplicity \pp collisions. Those intriguing results provide new insights into the origin of the flow-like signal in small collision systems.

%

\bibliographystyle{apsrev}
\typeout{}
\bibliography{TPC}

\end{document}